\documentclass[pre,showpacs,floatfix,reprint]{revtex4-1}

\usepackage{graphicx}
\usepackage{amsmath}
\usepackage{bm}
\usepackage{cases}

\DeclareMathOperator{\im}{Im}
\DeclareMathOperator{\re}{Re}
\newcommand{\qav}[1]{\langle {#1} \rangle}

\begin{document}
  \title{Quantum fluctuations and coherence in high-precision single-electron capture}
 \author{Vyacheslavs Kashcheyevs}
  \author{Janis Timoshenko}
  \affiliation{Faculty of Computing and Faculty of Physics and Mathematics, University of Latvia, LV-1586, Riga, Latvia}
\begin{abstract}
The phase of a single quantum state is undefined unless the history of its creation provides a reference
point. Thus quantum interference may seem hardly relevant for the design
of deterministic single-electron sources which strive to isolate individual charge carriers quickly and completely.
We provide a counterexample by analyzing the non-adiabatic
separation of a localized quantum state from a Fermi sea due to a closing tunnel barrier.
We identify the relevant energy scales and suggest ways to
separate  the contributions of
quantum non-adiabatic excitation and backtunneling to the rare non-capture events.
In the optimal regime of balanced decay and non-adiabaticity,
our simple electron trap turns into a single-lead
Landau-Zener-backtunneling interferometer, revealing the dynamical phase accumulated
between the particle capture and leakage.
The predicted ``quantum beats in backtunneling''
may turn the error of a single-electron source into a valuable signal
revealing essentially  non-adiabatic energy scales of a dynamic quantum dot.
\end{abstract}
\maketitle
Successful demonstration of electron-on-demand sources based on electrostatic modulation of nanoelectronic circuit elements such as dynamic
quantum dots \cite{blumenthal2007a,Giblin2012,kaestner2010d} or mesoscopic capacitors \cite{Feve2007,Feve2012exp} has offered a prospect of
building an electronic analog of few-photon quantum optics \cite{Zeilinger2012} that exploits the particle-wave duality and entanglement
of individual elementary excitations in a Fermi sea \cite{Bertoni2000,Olkhovskaya2008,SplettstoesserInterf2009,Haack2012}.
This ambitious goal is complemented by a long-standing challenge in quantum metrology \cite{Zimmerman2003} to untie the definition of ampere
from  the mechanical units of SI \cite{Milton2010} and implement a current standard
based on direct counting of discrete charge carriers.
So far the overlap between these research directions \cite{Brandes2010,fricke2011} has been rather limited arguably because
metrological applications strive to maximize the particle nature of on-demand excitations.
Optimizing the trade-off between speed and accuracy of single-electron isolation \cite{Giblin2012}
does require consideration of quantum error mechanisms such
as non-adiabatic excitation \cite{liu1993,Flensberg1999,kataoka2011} or backtunneling \cite{Aizin1998,fujiwara2008,kaestner2009a,kaestner2010a}. However,
these effects have been hard to
differentiate experimentally
owing to complexity of non-equilibrium many-particle
quantum dynamics \cite{Keeling2008} and experimental challenges in exercising
high-speed control of the electrostatic landscape. The quantum phase of
the captured particle has been considered thus far
as inconsequential for accuracy
and inaccessible for measurement unless the particle is ejected into a separate
interferometer \cite{Haack2012}.

In this work, we propose a new type of interferometry
to measure and thus control
the non-equilibrium energy scales governing
the decoupling of a dynamic quantum dot from a Fermi sea.
Remarkably, our approach requires neither multiple spatial paths \cite{SplettstoesserInterf2009,Haack2012} nor noise measurements \cite{Keeling2008,Olkhovskaya2008,Parmentier2012},
relying instead on quantum beats in spontaneous
emission of electrons back to the source lead.
Using a generic \cite{Jauho1994,Keeling2008} effective single-particle model
we predict an interference pattern in the charge capture probability
that reflects the dynamical quantum phase accumulated
between the isolation of the localized quantum state
and the onset of backtunelling.

\begin{figure}
 \includegraphics[width=\columnwidth]{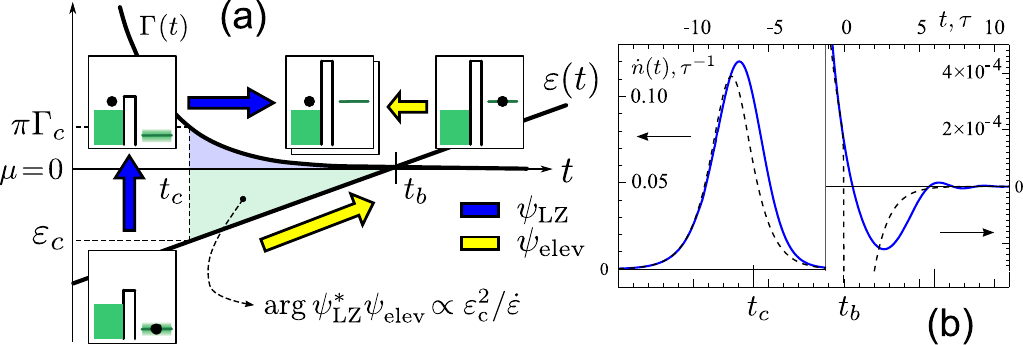}
  \caption{\label{fig:scheme}
  (color online) (a) Schematic plot of the rising level energy $\varepsilon(t)$ and exponentially
  decreasing broadening $\Gamma(t)$. Pictograms illustrate the two paths for interference:
  dark (blue) arrows mark non-adiabatic excitation followed by phase accumulation
  above the Fermi edge $\mu$, light (yellow) arrows show adiabatic ``elevator ride''
  and backtunneling. (b) Instantaneous current flowing into the quantum dot, $\dot{n}(t)$.
  Thick (blue) line -- exact Eq.~\eqref{Flensberg7},
  thin dashed line --  quantum-broadened Markov approximation, Eq.~\eqref{eq:labelKinetic}.
  Model parameters: $\varepsilon_c/\Delta_{\text{ptb}}\!=\!-6$,
  $\Gamma_c/\Delta_{\text{ptb}}\!=\!1.2$, $T=0$.
  }
\end{figure}

Our innovations can be explained within
a simple two-path picture of Mach-Zehnder interferometry in time domain,
see Fig.~\ref{fig:scheme}(a).
The first branch is quantum excitation above the Fermi edge in the source lead due to
finite time-scale $\tau$ for the pinch-off of tunneling \cite{Flensberg1999} [see dark (blue) arrows in  Fig.~\ref{fig:scheme}(a)].
The corresponding energy spread and
the (small) path-splitting amplitude can be estimated by the energy-time uncertainty, $\Gamma_c \equiv \hbar/(\pi \tau)$, and
the Landau-Zener theory \cite{LandauQMlz}, respectively.
The second branch is  adiabatic lifting of the occupied energy level followed by splitting of
a small amplitude back into the lead once the level emerges above the Fermi sea
[see light (yellow) arrows in Fig.~\ref{fig:scheme}(a)].
This elevation-backtunneling branch is sensitive
to  electrostatic cross-coupling between the barrier and the bottom of the confining potential well;
we quantify this crosstalk by \emph{plunger-to-barrier} ratio $\Delta_{\rm ptb} \equiv \dot{\varepsilon} \tau$
(shift of the localized energy level $\varepsilon$ during the characteristic decoupling time).
The output ports of the interferometer can be read either by detecting excitations
created in the Fermi sea, or, more conveniently, by measuring
the charge capture probability.
The latter is accessible in experiment by repeated ejection
of the captured electrons into a collector lead and measuring
the resulting dc current~\cite{blumenthal2007a,fricke2011,Giblin2012,fujiwara2008,kaestner2009a,kaestner2010a}.
 The beam-splitters are tuned by $\Delta_{\rm ptb}/\Gamma_c$ to maximize contrast,  the phase measurement reveals the ``elevator speed''
$\dot{\varepsilon} \propto \Delta_{\rm ptb} \Gamma_c $, and the temperature smearing allows absolute calibration of the energy scales.
Our scheme is conceptually related to
Landau-Zener-St\"uckelberg interferometry
\cite{LevitovScience2005,Kraemer2007,Shevchenko2010} which measures the relative
dynamical phase of  discrete states via creation of superposition in sequential
non-adiabatic level crossings.
In contrast, we propose to access the phase of a \emph{single} localized state
measured against a reference point in the continuum
(defined by a sufficiently fast decoupling
and a sharp Fermi edge).

The proposed measurement addresses
a persistent challenge in robust utilization of electrostatically defined
quantum dot devices: control of rate ($\tau$), type (barrier versus plunger) and magnitude ($\Delta_{\rm ptb}$) at which
external voltage pulses are converted into the time-dependent potential guiding individual transport electrons on the chip.
Although parametric time-dependency of the electronic matrix elements is the standard input for theory \cite{Jauho1994,moskalets2002B},
in practice mesoscopic fluctuations (due to distribution and charge switching  dynamics of impurities,
finite fabrication precision etc.) and challenges of signal propagation at high frequencies (GHz range)
often require measuring the characteristic quantities on sample-to-sample basis.
For near-equilibrium, bias spectroscopy \cite{kouwenhoven2001} provides a versatile tool, but for the large-amplitude, high-frequency modulation
the options are limited~\cite{blumenthal2007a}.
As a foreseeable direct application of our results we expect
a reliable measurement of $\Gamma_c/\Delta_{\text{ptb}}$ to help settling the debate on the
fundamental factors limiting the precision
of the state-of-the-art single-electron-based current sources~\cite{Giblin2012}.
More generally, we hope our analysis
will facilitate the crossover of ideas between fundamental and applied
directions of single electronics.

\paragraph*{Model and formalism.}
The model is described by an effective single-particle Hamiltonian  $\mathcal{H}=  \varepsilon(t) d^{\dagger} d^{}+
\sum_k \{ \epsilon_k^{} c_{k}^{\dag} c_{k}^{} + V(t) [  c_k^{\dag}  d^{}_{} + d^{\dagger}_{} c_k^{} ] \} $,
where $d^{\dagger}_{}$ creates a localized non-degenerate electronic state in the dot and $c_{k}^{\dagger}$ --- a quasi-continuous state
in the lead.
The lead
is connected to a thermal reservoir with chemical potential $\mu\!=\!0$ and temperature $T$.
Employing a time-dependent tunneling Hamiltonian
relies on timescale separation \cite{Jauho1994}: fast screening
in the leads determines the instantaneous values
of the slowly  varying   parameters for
the underscreened region (the quantum dot).
We choose  $k$-independent real $V$ and the wide-band limit so that
$\Gamma(t) \equiv 2 \pi \rho V^2(t)$ and $\varepsilon(t)$ are the fully
dressed elastic width and the on-site energy respectively ($\rho$ is the density of states in the lead).

The quantum kinetic equation for the average occupation within the model, $n(t) \equiv \qav{d^{\dagger}_{}(t) d^{}(t)}$,
is given by the non-equilibrium Green functions theory \cite{Jauho1994},
\begin{multline}
\label{Jauho:1b}
 \hbar \dot{n}(t) =  - \Gamma(t) n(t) -
 \\ - \int \frac{f(\epsilon) }{\pi \hbar} \im\!
 \int\nolimits_{- \infty}^{t} \!\!\!\! \sqrt{ \Gamma(t) \Gamma(t')} G(t,t') e^{i \epsilon (t-t')/\hbar} d t' d \epsilon \, ,
\end{multline}
where $f(\epsilon)$ is the Fermi distribution,
and $G(t,t')$ is the retarded Green function of the level,
\begin{equation}\label{FJauho:3mod}
G(t, t') = -i \Theta(t-t')  e^{-i \int \nolimits_{t'}^t d t_1 [ \varepsilon(t_1) - i \Gamma(t_1)/2] / \hbar } \, .
\end{equation}

Integrating Eq.~\eqref{Jauho:1b} gives [cf.\ Eq.~(44) of Ref.~\onlinecite{Jauho1994}]
\begin{align}\label{Flensberg7} \footnotesize
    n (t) & =  \int \!\frac{f(\epsilon) }{2 \pi \hbar^2}
    \Bigl | \int\nolimits_{- \infty}^{t}\!\!\!\!\! d t' \sqrt{\Gamma(t')}  G(t, t')   e^{- i \epsilon t'/\hbar}    \Bigr |^2 d \epsilon \, .
\end{align}
Equation \eqref{Flensberg7} is the sum of probabilities to be scattered into the localized state from an occupied state in the continuum;
one can show \cite{hanggiShow,Arrachea2006} (see Appendix \ref{app:A}) 
that it agrees with the Floquet formalism \cite{hanggiFloquet,moskalets2002B} which
is often used in the scattering form to study
single-charge emitters \cite{Floquet2008prl,battista2011,Parmentier2012}.
In this work we consider electron trapping achieved
by reducing  $\Gamma(t)$ to zero as $t\to \infty$ and compute
the capture probability $n_f \equiv \lim_{t\to\infty} n(t)$.

The history of parametric time-dependence
that affects  $n(t)$ is limited by a finite memory time $\tau_{\text{mem}} \equiv \hbar \text{min} \{(kT)^{-1}, \Gamma^{-1}(t)\}$ due to (a) lead-induced dephasing of the discrete state,
and (b) thermal smearing in the reservoir.
One can show~(see Appendix \ref{app:B})
that for sufficiently slow processes, when  $| \dot{\varepsilon} | \tau_{\text{mem}}^2 \ll \hbar$ (no phase rotation) and
$|\dot{\Gamma}| \tau_{\text{mem}} \ll \Gamma$ (well-defined $\Gamma$),
the exact kinetic equation \eqref{Jauho:1b} can be replaced by a (quantum broadened) Markov approximation,
\begin{align} \label{eq:labelKinetic}
    \hbar \dot{n}(t) & = - \Gamma(t) \left \{ n(t) - n_{\text{eq}}\left [\varepsilon(t), \Gamma(t)\right]
    \right \} \, ,
\end{align}
with parametrically defined standard equilibrium occupation,
$ n_{\text{eq}}(\varepsilon, \Gamma) =(2 \pi)^{-1} \int d \omega\,\Gamma   f(\omega) /[(\omega-\varepsilon)^2+\Gamma^2/4]$.
Although Eq.~\eqref{eq:labelKinetic} still permits strongly non-adiabatic scenarios \cite{Kaestner2007c}, 
for the decoupling problem at hand
$\tau_{\text{mem}}$ diverges at zero temperature as $t\to \infty$, thus we must use Eq.~\eqref{Flensberg7}.

We define a crossover moment $t_c$ as the earliest time from which the quantum phase of the level can be preserved,
$\int_{t_c}^{+\infty}\!\Gamma(t) dt\!=\!\hbar$,
and explore exponential
time-dependence \cite{Flensberg1999}
of $\Gamma(t)$ around $t_c$,
\begin{align} \label{eq:parametrziedGamma}
  \Gamma(t) = \pi \Gamma_c e^{-(t-t_c)/\tau} \, ,
\end{align}
accompanied by a linear shift of energy \cite{Keeling2008,Kaestner2007c,fujiwara2008,kaestner2009a},
\begin{align} \label{eq:epsilon}
  \varepsilon(t) =\varepsilon_c +\Delta_{\text{ptb}}(t-t_c)/\tau \, .
\end{align}
The shape of $\varepsilon(t)$ and $\Gamma(t)$ for $t \ll t_c $
as well as the initial conditions for Eq.~\eqref{Jauho:1b} are  irrelevant for
$n_f$ if the ansatz \eqref{eq:parametrziedGamma}
holds from a few $\tau$ before $t_c$.
Whether the last particle exchange between the dot and the source (most likely to occur around $t \approx t_c$)
results in a captured electron ($n_f \approx 1$) or a hole
($n_f \approx 0$) depends on the position of
$\varepsilon_c$
with respect to $\mu$ (below or above, respectively).

\paragraph*{Experimental realization.}
A prototypical realization for the model is a small near-empty electrostatically defined quantum
dot with large level spacing and charging energy, as in, e.g.
\cite{kaestner2010d,kataoka2011}. A linear ramp of voltage
$\mathcal{V}_1(t)$ on a gate that defines the tunnel barrier between the source and the quantum dot
creates the time-dependencies \eqref{eq:parametrziedGamma} and \eqref{eq:epsilon}
while a static voltage $\mathcal{V}_2$ on another gate can be used to tune the decoupling energy,
$\varepsilon_c =-\alpha \mathcal{V}_2 \Delta_{\text{ptb}}+\text{const}$.
$\Gamma_c \sim \tau^{-1}$ is inversely proportional to the rise-time of the $\mathcal{V}_1(t)$ pulse.

Once $\Gamma(t)$ becomes negligible ($t\to \infty$ in our model),
further modulation of the confining potential can ensure complete ejection of the captured  electrons into the drain lead \cite{Giblin2012,kaestner2010a}.
Repeating the whole cycle with a frequency $f \ll \tau^{-1}$ and measuring the dc component of the source-drain pumping
current $I(\mathcal{V}_2)$ can provide accurate data on $n_f (\varepsilon_c)=I/(e f)$ (here  $e$ is the electron charge).
Note that
$\alpha$  is the fitting parameter \cite{kaestner2009a,Giblin2010,Giblin2012} of the decay cascade model \cite{kaestner2010a},
see Eq.~\eqref{eq:limitDoubleExp} below.

\paragraph*{Qualitative picture.}
The essential features of the model
can be seen in the time-dependent ensemble-average current $\dot{n}(t)$ for $\Delta_{\text{ptb}} \approx \Gamma_c$
and $\epsilon_c < -\Delta_{\text{ptb}}$  at low temperature,
see Fig.~\ref{fig:scheme}(b).
Before the formation of the localized level, at times $t \ll t_c-\tau \ln |\epsilon_c| /\Gamma_c$,
the average occupation number remains constant and equal to the average density of electrons per quantum state in the lead
($1/2$ in our dispersionless model). A large peak in the current near $t_c$ marks adiabatic filling of the rapidly narrowing level.
Much smaller
opposite sign
feature
indicates
the onset of backtunneling at $t>t_b$ with $t_b$ defined by $\varepsilon(t_b)\!=\!\mu$.
It is instructive to contrast the exact $\dot{n}(t)$ with the solution of the Markov equation \eqref{eq:labelKinetic} shown by the dashed line
in Fig.~\ref{fig:scheme}(b).
The level of agreement correlates with
the condition $\Gamma(t) > \Gamma_c$ being fulfilled exponentially well for $t \ll t_c$, breaking down around $t_c$, and being strongly violated
for $t>t_b$.

\paragraph*{Probability of charge capture.}
Our main result is
\begin{align}
  n_f  = & \int \frac{d \epsilon}{2 \pi^2 \Gamma_c} f(\epsilon)    \label{eq:mainresult} \\
  & \times \Bigl \lvert \int_{-\infty}^{\infty} \!\!\! \exp \Bigl [-\frac{x+e^{-x}}{2} +i \frac{\Delta_{\text{ptb}}}{2 \pi \Gamma_c} \bigl( x -\frac{\epsilon-\varepsilon_c}{\Delta_{\text{ptb}}}
  \bigr)^2
   \Bigr ] dx \Bigr \rvert^2 \, . \nonumber
 \end{align}
$n_f(\varepsilon_c)$ is a step-like function changing from $1$ to $0$ as $\epsilon_c$ goes from $-\infty$ to $+\infty$.
The limit forms are
\begin{subnumcases}{n_f\!=\!}
    \!\! f(\varepsilon_c) 
    ,                                               & \!\!\!\! $\Delta_{\text{ptb}}, \Gamma_c\!\to\!0 $ \label{eq:limitFermi}
     \\
    \!\! (2/\pi) \tan^{-1} e^{-\varepsilon_c/\Gamma_c}  , & \!\!\!\! $kT, \Delta_{\text{ptb}}\!\to\!0$  \label{eq:limitFlensberg}
    \\
    \!\! e^{-e^{\varepsilon_c/\Delta_{\text{ptb}}}}  . & \!\!\!\! $kT, \Gamma_c\!\to\!0$ \label{eq:limitDoubleExp}
  \end{subnumcases}
The limit $\eqref{eq:limitFermi}$ corresponds to a sudden decoupling from equilibrium at $\varepsilon(t_c)=\varepsilon_c$,
the corresponding step is symmetric under $\varepsilon_c \to -\varepsilon_c$ and maps out the thermal distribution
\cite{yamahata2011}. The limit \eqref{eq:limitFlensberg}
reproduces the result of Flensberg \emph{et al.} \cite{Flensberg1999}, a symmetric step of width $\Gamma_c$.
The double-exponential shape \eqref{eq:limitDoubleExp} has been
predicted previously \cite{kaestner2009a,fujiwara2008,kaestner2010a} from a master equation
and validated experimentally \cite{Giblin2010,Giblin2012} with up to $10^{-6}$ relative accuracy.
Our derivation reveals the plunger-to-barrier ratio $\Delta_{\text{ptb}}$ as the energetic measure of the step width.

\begin{figure}
  \includegraphics[width=0.9\columnwidth]{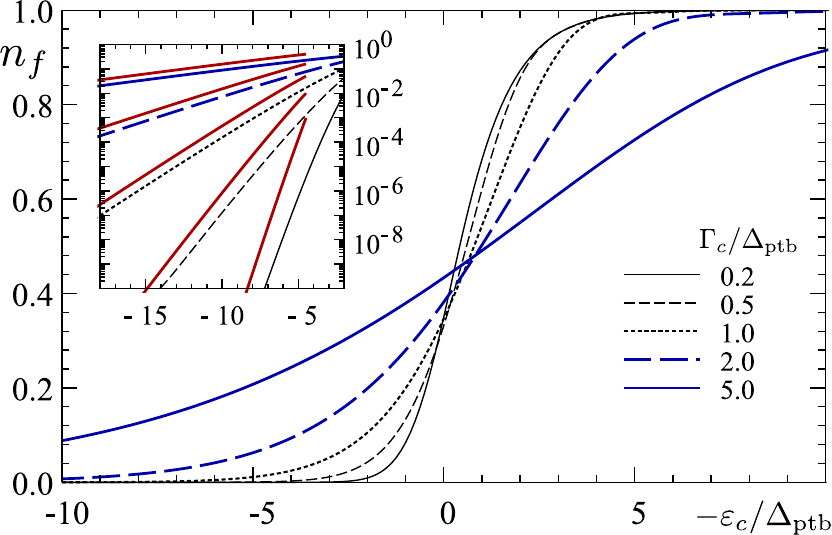}
  \caption{\label{fig:nf} (color online)
   Charge capture probability $n_f$ as a function of the level energy $\varepsilon_c$ at the decoupling moment.
   $T=0$ and $\Gamma_c/ \Delta_{\text{ptb}}$ is varied as indicated. The inset show the same quantity in the logarithmic scale
   with additional thick (red) lines marking the asymptotics \eqref{eq:limitleft}.
  }
\end{figure}

Finite temperature is accounted for by simple thermal smearing,
$n_f(\varepsilon_c) = \int n_f(\varepsilon_c-\epsilon)\rvert_{T=0}  (-\partial f/\partial \epsilon)  d \epsilon$,
thus we focus mainly on $T=0$.
Figure \ref{fig:nf} shows evolution of the lineshape $n_f(\epsilon_c)$ as $\Gamma_c/\Delta_{\text{ptb}}$ is increased.
The asymmetry
around $\varepsilon_c =0$ gets ``inverted'' with respect to the double-exponential \eqref{eq:limitDoubleExp} at $\Gamma_{c}\!=\! (2\!\ldots \!3) \Delta_{\text{ptb}}$
before approaching the symmetric limit   \eqref{eq:limitFlensberg}.
Non-perturbative asymptotics of Eq.~\eqref{eq:mainresult}
for $\epsilon_c \gg \max [ \Delta_{\text{ptb}}, \Gamma_c ]$
gives an exponential dependence with a power-law prefactor,
\begin{align} \label{eq:limitleft}
 n_f \sim  \frac{2}{\pi} \left (\frac{2 \epsilon_c}{\pi \Gamma_c} \right )^{\Delta_{\text{ptb}}/\Gamma_c} e^{-\varepsilon_c/\Gamma_c}
 \, , \quad  T \to 0 \, ,
\end{align}
shown in the log-scale inset of Fig.~\ref{fig:nf}.
This is similar to the Fermi function tail, $\sim e^{-\varepsilon_c/kT}$, which dominates the small-$n_f$ asymptotics
for $k T > \Gamma_c$ regardless of $\Delta_{\text{ptb}}$.
Thus we conclude that
both quantum and thermal fluctuations always  trump
the double-exponentially suppressed backtunneling for small $n_f$.
This finding may have implications for the minimal slope position on the plateaux between successive
current quantization steps in single-gate pumps~\cite{kaestner2009a,Giblin2010}.
A change in slope of $\log n_f$ versus $\varepsilon_c$ as the barrier-closing timescale $\tau$ is reduced
may signal that $\Gamma_c / kT \ge 1$, although this effect on its own would be hard to differentiate from local heating.

\begin{figure}
  \includegraphics[width=\columnwidth]{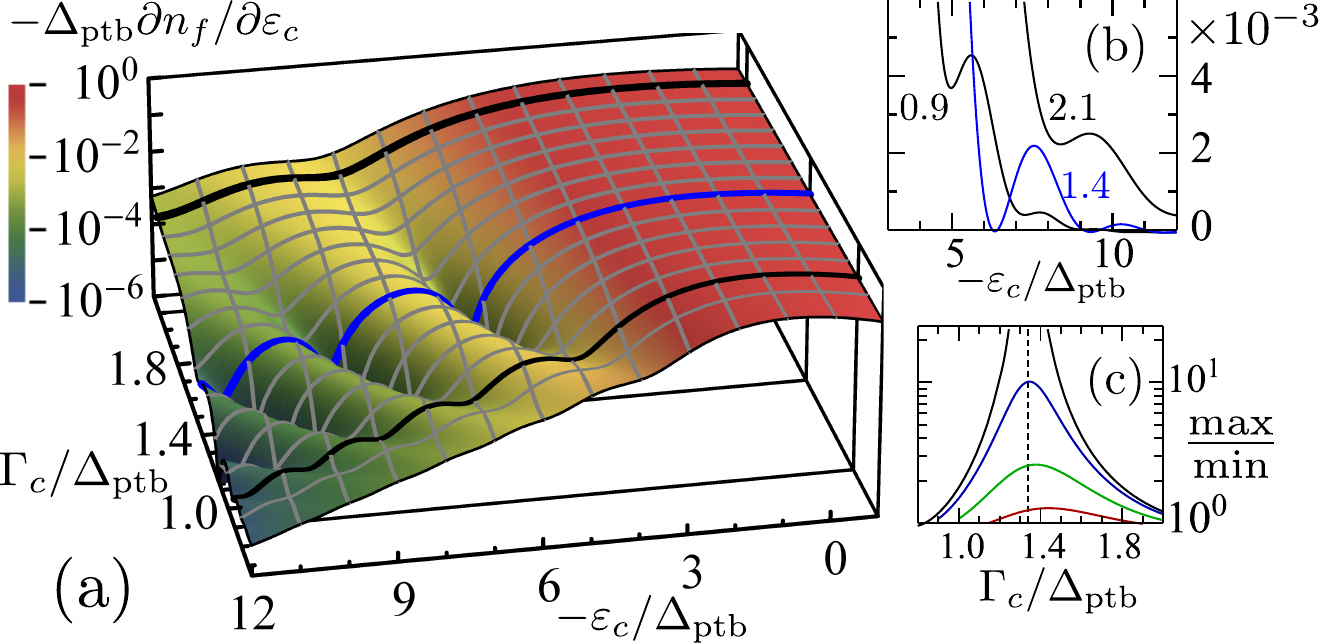}
  \caption{\label{fig:ripples}
   (color online)  
   (a) The derivative  $-\Delta_{\text{ptb}} \partial n_f/\partial \varepsilon_c$ (vertical axis, log scale)
   as a function of the decoupling energy $\varepsilon_c$ and the ratio
    $\Gamma_c/\Delta_{\text{ptb}}$ at $T=0$. 
   Three cuts at $\Gamma_c/\Delta_{\text{ptb}}=0.9, 1.4$ and $2.1$ are shown on a linear scale in (b).
   (c) The ratio of the first interference
   maximum in $-\partial n_f(\varepsilon_c)/\partial \varepsilon_c$  to the first minimum  on a  log scale
   as a function of $\Gamma_c /\Delta_{\text{ptb}}$
   as $k T / \Delta_{\text{ptb}}$ increases through $0, 0.1, 0.2$ and $0.3$ (from the topmost to the lowermost curve).
    }
\end{figure}

\paragraph*{Quantum oscillations.}
The asymptotics of $1\!-\!n_f$ at large negative $\varepsilon_c$ switches from $\sim e^{\varepsilon_c/\Delta_{\text{ptb}}}$ to
$\sim e^{\varepsilon_c/\Gamma_c}$ at $\Gamma_c=\Delta_{\text{ptb}}$ via a surprising sequence of miniplateaus.
The latter can be seen as ripples in the derivative $\partial n_f/\partial \varepsilon_c$, as shown in Fig.~\ref{fig:ripples}(a)-(b).
Although the oscillation amplitude decays exponentially, the non-monotonic behavior of $\partial n_f(\varepsilon_c)/\partial \varepsilon_c$
is manifest in the range of $\Gamma_c\!=\!(0.8 \ldots 2.2)\Delta_{\text{ptb}}$ at temperatures up to $kT \approx 0.35 \Delta_{\text{ptb}}$, see
Fig.~\ref{fig:ripples}(c).

We interpret the oscillations as interference between two excitation paths that promote lead electrons above the Fermi energy.
One path is capture-elevation-backtunneling similar to elevator resonance activation proposed by Azbel'~\cite{Azbel1992}.
As seen from the scattering interpretation of Eq.~\eqref{eq:mainresult}, the electrons most likely to be captured
have incoming energies $\epsilon \approx \varepsilon_c$, thus  we estimate the amplitude to be raised by an ``elevator ride''
from $\varepsilon_c $ to an energy $\epsilon_e > \mu$  by a three-amplitude product,
\begin{align}
\psi_{\text{elev}} (\epsilon_e) \propto   V(t_c) e^{-i \int_{t_c}^{t_e} \varepsilon(t) d t /\hbar} V(t_e)  \, ,
\end{align}
where the exit time $t_e>t_b$ is determined from $\varepsilon(t_e)=\epsilon_e$.

The other path from $\epsilon \approx \varepsilon_c$ to $\epsilon = \epsilon_e$ is Landau-Zener-like excitation
 due to time-dependence of $\Gamma(t)$. We estimate the corresponding amplitude by following the Landau solution
 of  a two-level problem  \cite{LandauQMlz}  with  matrix  elements $H_{11}\!=\!\epsilon_e$, $H_{22}\!=\!\varepsilon_c$,
 and $H_{12}\!=\!H_{21}\!=\!\Gamma(t)$~\cite{Note2}.
 The adiabatic eigenvalues $E_1(t)$, $E_2(t)$ have a gap
 $\Delta E(t) =\sqrt{4 \Gamma^2(t)\!+\!(\epsilon_e\!-\!\varepsilon_c)^2}>0$ on the real axis
 but become degenerate
 if analytically continued into the complex $t$-plane.
The branching point of $E(t)$ with the smallest positive imaginary part,
 $t_0\!=\!t_c\!+\!\tau \ln [2 \Gamma_c/(\epsilon_e\!-\!\varepsilon_c)]\!+\!i \pi \tau  /2$, determines
 the transition probability \cite{LandauQMlz}  $\exp[ -2 \im \int_{\re t_0}^{t_0}\!\Delta E(t) d t /\hbar ]\!=\!\exp [-\pi \tau (\epsilon_e\!-\!\varepsilon_c)/\hbar]$.
 Neglecting  pre-exponential and logarithmic terms,
 the  amplitude for excitation at $t_c$  and evolution up to $t_e$ is
\begin{align}
\psi_{\text{LZ}} (\epsilon_e) \propto   e^{-(\epsilon_e-\varepsilon_c)/(2 \Gamma_c)-i \varphi_0} e^{-i \epsilon_e  (t_e -t_c) / \hbar} \, ,
\end{align}
where $\varphi_0$ is the phase of the Landau-Zener transition (Stokes phase) known to depend weakly on energy \cite{Shevchenko2010}.

\begin{figure}
  \includegraphics[width=0.95\columnwidth]{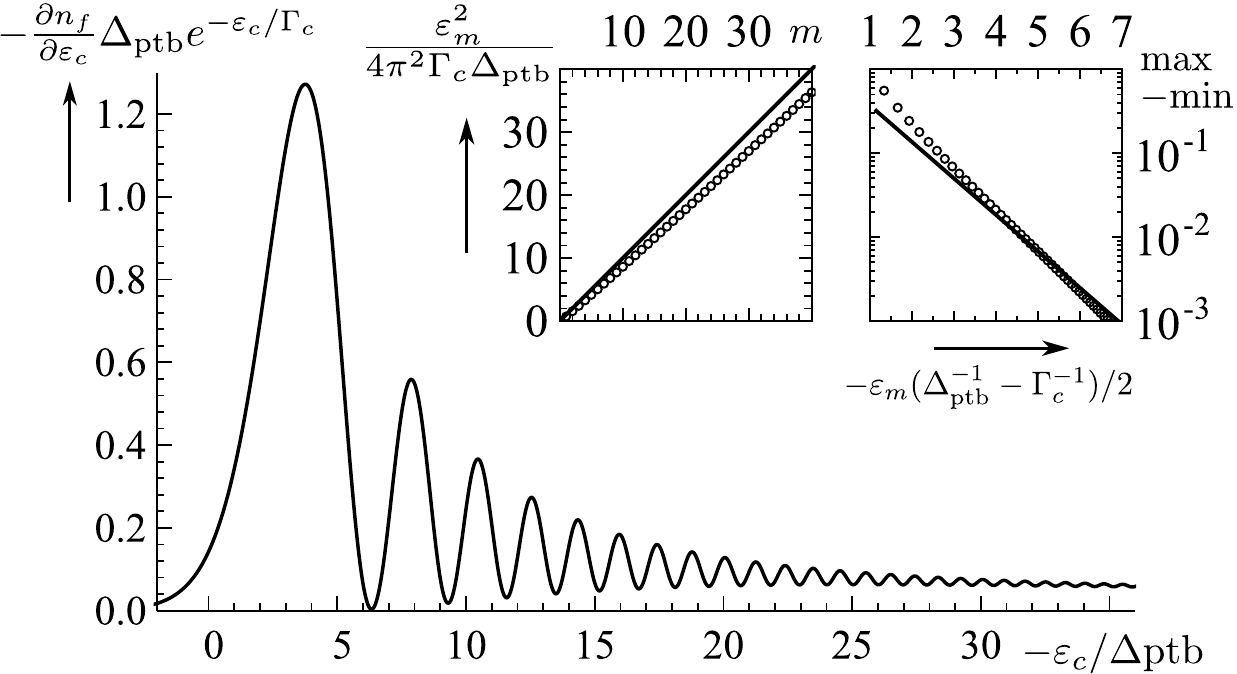}
  \caption{\label{fig:oscproof}
   Main panel: the derivative  $-\partial n_f/\partial \varepsilon_c$ multiplied by
   $\Delta_{\text{ptb}}  e^{-\varepsilon_c/\Gamma_c}$ for $\Gamma_c/\Delta_{\text{ptb}}=1.4$ and $T\!=\!0$.
   Left inset: the positions $\varepsilon_m$ of the minima in the main graph versus their index number, $m$, scaled according
   to the two-path formula \eqref{eq:twoslit}.
   Right inset: the absolute difference between a maximum  and the preceding minimum of the oscillations shown in the main panel
   on a log scale versus a scaled $-\varepsilon_m$.
   The straight line of slope of $\pm 1$ in both insets is plotted according to Eq.~\eqref{eq:twoslit}.}
   \end{figure}

We estimate the total non-capture probability as
\begin{align}
  1-n_f \propto \int_{\mu=0}^{\infty} \!\!\!\! d\epsilon_e | \psi_{\text{LZ}}(\epsilon_e)+ \psi_{\text{elev}}(\epsilon_e) |^2.
\end{align}
The competition of $ | \psi_{\text{LZ}}|^2$ and $|\psi_{\text{elev}} |^2$ at large negative $\varepsilon_c$ agrees with the asymptotic envelope of $n_f$
while $\re \psi_{\text{LZ}}^{}(0) \psi_{\text{elev}}^{\ast}(0)$  gives the oscillating part of
$\partial n_f/\partial \varepsilon_c$,
\begin{align} \label{eq:twoslit}
e^{\varepsilon_c (\Delta_{\text{ptb}}^{-1}+\Gamma_c^{-1}) / 2} \cos \Bigl (\varphi_0\!+\!\frac{ \epsilon_c^{2}}{2 \pi \Gamma_c \Delta_{\text{ptb}}} \Bigr ) \,.
\end{align}
The argument of the cosine is the dynamical phase accumulated between $t_c$ and $t_b$, see the shaded triangle in Fig.~\ref{fig:scheme}.
Despite the crudeness of approximations leading
to Eq.~\eqref{eq:twoslit}, the result agrees well with the exact solution Eq.~\eqref{eq:mainresult} both in amplitude and phase, as shown in Fig.~\ref{fig:oscproof}.

\paragraph*{Feasibility.}
The oscillation effect does not rely on any finetuning (apart from
balancing the interferometer, $\Delta_{\text{ptb}} \approx \Gamma_c$) thus it should be robust
against deviations from the assumed time-dependencies. 
The most important foreseeable limitation is the overlap of additional interference modes
which must be separated by a sufficiently large on-the-dot level spacing $\Delta \varepsilon$
from $t \approx t_c$ onward; our single-mode formula \eqref{eq:mainresult} is limited to $-\varepsilon_c < \Delta \varepsilon$ requiring
$\Delta \varepsilon \gtrsim 8 \Delta_{\text{ptb}}$ to resolve the first interference minimum (see Appendix \ref{app:C}).
Using a recent report \cite{kataoka2011}
on out-of-equilibrium excited states in a rapidly decoupling dynamic quantum dot
we read $\alpha= 0.28 \,\rm{mV}^{-1}$ from  100 MHz data in Fig.~1 of Ref.~\onlinecite{kataoka2011}
and
estimate the gap to the first exited state observed at 1 GHz 
to be
$\Delta \varepsilon / \Delta_{\text{ptb}}  \approx 4.5$ which is the same order of magnitude as our requirements. 
Measurements of a similar device \cite{Giblin2012} have just reached the required accuracy threshold
for $n_f(\varepsilon_c)$
suggesting the proposed dynamical phase interferometry is feasible with current technology.

\paragraph*{Conclusions.} We have considered the quantum dynamics of isolating a single particle
from a Fermi sea by a closing tunnel barrier and proposed ways to measure
the relevant non-equilibrium energy scales.
Time-domain interferometry
revealing the dynamical phase of the confined particle offers a unexpected ``built-in''
instrument  to quantify and separate the quantum effects hampering deterministic electron-on-demand operation
which is a cornerstone for future on-the-chip electron quantum optics
and a quantum realization of the ampere.

\acknowledgments
We thank Bernd Kaestner for discussions. This work has been supported by
ESF  project No.~2009/0216/1DP/1.1.1.2.0/09/APIA/VIAA/044.

\appendix
\renewcommand{\theequation}{A\arabic{equation}}%
\renewcommand{\thefigure}{A\arabic{figure}}%
\setcounter{equation}{0}
\setcounter{figure}{0}
\section{Relation to Floquet formalism\label{app:A}}
Here we demonstrate the connection of Eqs.~(1) and (3) in our paper
to the instantaneous current formulae derived in Ref.~\onlinecite{Parmentier2012} for a periodically modulated
mesoscopic system.
Using Eqs.~(9) and (10) of Ref.~\onlinecite{Parmentier2012}, the average
incoming current can be expressed as
\begin{align}\label{eq:Ifloquet}
  \qav{\hat{I}(t)} = \frac{e}{2 \pi \hbar} \int d \epsilon f(\epsilon)
  \Bigl (
    \bigl \lvert \sum_m e^{i m \Omega t} U_m(\epsilon-m \hbar \Omega) \bigr \rvert^2 -1
  \Bigr ) \, ,
\end{align}
where $2 \pi /\Omega$ is the period of parameter modulation. The Floquet scattering matrix $U_m(\epsilon)$ is a double-Fourier representation
of the time evolution operator for the lead electrons [see Eqs.\ (5)-(6) of Ref.~\onlinecite{Parmentier2012}],
\begin{align} \label{eq:Floqdef}
  U(t,t') & = \frac{1}{2 \pi}  \sum_m \int d \omega e^{-i \omega (t-t')+i m \Omega t'} U_m(\hbar \omega) \, .
\end{align}
On the other hand, scattering amplitude in the lead and the retarded Green
function on the quantum dot are related by the standard equation
\begin{align} \label{eq:FisherLee}
  U(t,t') = \delta (t-t') - 2 \pi i \rho V(t) G(t,t') V(t')/\hbar \, ,
\end{align}
which follows directly form the equation of motion for the Heisenberg operators in the leads.
Using Eqs.~\eqref{eq:Floqdef} and \eqref{eq:FisherLee} to expresses the
time-dependent $\mathcal{T}$-matrix
\begin{align}
    \mathcal{T}(\epsilon,t) & \equiv {\hbar^{-1}}  \sqrt{\Gamma(t)} \int_{-\infty}^{+\infty} \nonumber
  \!\!\!\!  G(t,t') \sqrt{\Gamma(t')} \, e^{i \epsilon (t-t') /\hbar} dt' \\
  & =  -i + i \sum_m e^{i m \Omega t} U_m(\epsilon-m \hbar \Omega) \, ,
\end{align}
allows re-writing Eq.~\eqref{eq:Ifloquet} as
\begin{align} \label{eq:floq2} \tag{\ref{eq:Ifloquet}$'$}
  \qav{\hat{I}(t)} =\frac{e}{2 \pi \hbar} \int d \epsilon \, f(\epsilon) \left [ |\mathcal{T}(\epsilon,t)|^2 + 2 \im \mathcal{T}(\epsilon,t) \right ]  \, .
\end{align}
Since our result (3) for the time-dependent level occupation can be written as
\begin{align}
  \Gamma(t) n(t)=  \frac{1}{2 \pi} \int f(\epsilon) |\mathcal{T}(\epsilon,t)|^2  d \epsilon
  \tag{3$'$} \, ,
\end{align}
equivalence between the Floquet current \eqref{eq:Ifloquet} derived in Ref.~\onlinecite{Parmentier2012}
and the quantum kinetic equation (1) from the non-equilibrium Green functions theory is now explicit,
$\qav{\hat{I}(t)} = -e \dot{n}(t)$.

\section{Markov approximation\label{app:B}}
The role of quantum effects becomes clearer if we contrast Eq.~\eqref{Jauho:1b}
with a Markov master equation.
Performing the frequency integral in Eq.~(1) 
one gets
\begin{multline}
\label{Jauho:1bTransf} \tag{1$'$} 
 \hbar \dot{n} =  - \Gamma(t) n(t) + 2 \re \int_{-\infty}^{t} \! dt' \sqrt{ \Gamma(t) \Gamma(t')}  \tilde{f}^{\ast}(t-t')  \\
 \times  \exp \left \{-i \int_{t'}^t \! d \tilde{t} \, [ \varepsilon(\tilde{t}) - i \Gamma(\tilde{t})/2] / \hbar \right \}
  \, ,
\end{multline}
where $\tilde{f}(t)=\delta(t)/2+ i kT /[2 \hbar \sinh ( \pi kT t/\hbar)]$.
The main contribution to
the memory  integral in Eq.~\eqref{Jauho:1bTransf} is from an interval
of length
$\tau_{\text{mem}} =\hbar \text{min} \{ (kT)^{-1}  , \Gamma^{-1}(t) \}$.
If the time-dependence of $\varepsilon(t')$  and $\Gamma(t')$ can be ignored
around $t\approx t'$ over a time span longer than $\tau_{\text{mem}}$  then
the exponential in Eq.~\eqref{Jauho:1bTransf} can be approximated by
$e^{- i [\epsilon(t) (t-t')+\Gamma(t) |t-t'|/2] /\hbar}$ and
the memory integral evaluated as if the time-dependences were frozen.
This procedure results in the quantum broadened Markov approximation, see Eq.~\eqref{eq:labelKinetic} of the main text. %
The necessary conditions
are $| \dot{\varepsilon} | \tau_{\text{mem}}^2 \ll \hbar$ (no phase rotation) and
$|\dot{\Gamma}| \tau_{\text{mem}} \ll \Gamma$ (well-defined $\Gamma$), or, for our special  model,
\begin{align} \label{eq:kineticConditions}
  \Gamma_c \, , \; (\Gamma_c \Delta_{\text{ptb}})^{1/2}  & \ll \max [ \Gamma(t), k T ]  \, .
\end{align}
Thus the Markov approximation \eqref{eq:labelKinetic}
is adequate
when the quantum broadening due to time-dependence of rates ($\hbar/\tau$)  and
energy ($\sqrt{\hbar |\dot{\varepsilon}|}$) is smaller than the quasistatic level width (thermal, $k T$, or tunneling, $\Gamma$).

\section{Effect of excited states\label{app:C}}
\begin{figure*}[!t]
  \begin{center}
    \includegraphics[width=0.8\textwidth]{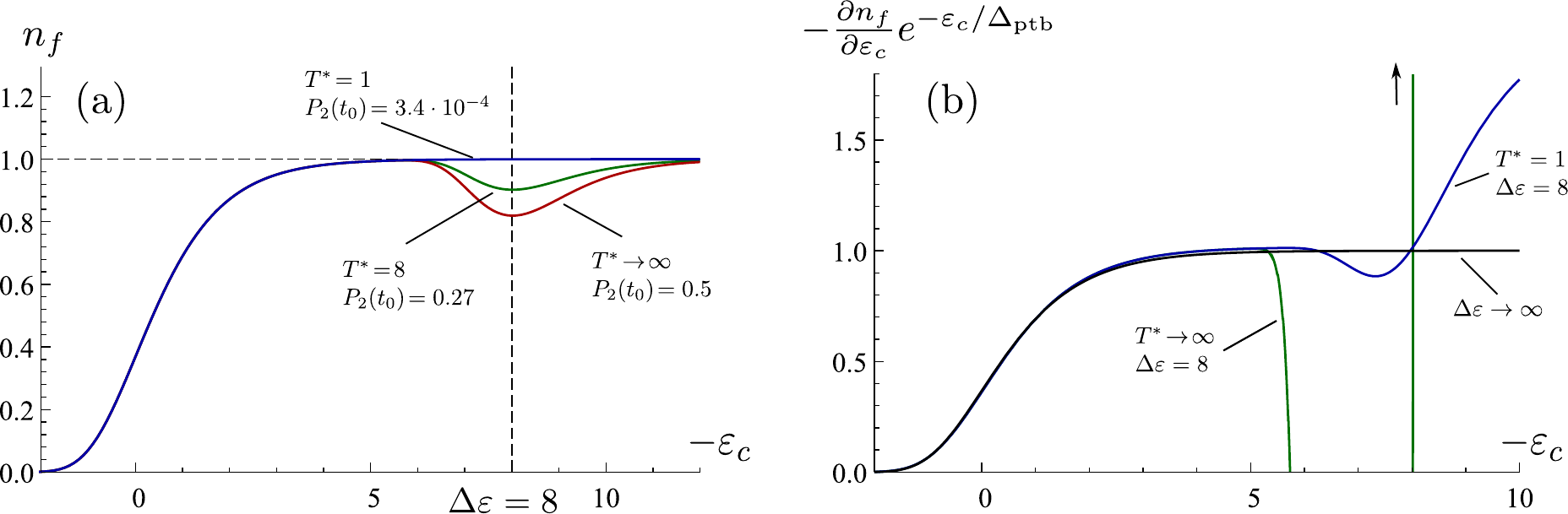}
  \end{center}
  \caption{Charge capture probability computed within a two-level model defined by Eqs.~\eqref{eq:parametrziedGamma}, \eqref{eq:alleps2},
  and \eqref{eq:mastereq}. All energies are in units of $\Delta_{\rm ptb}$ and $\Delta_{\rm ptb} \gg T\to 0$. \label{fig:S1}}
\end{figure*}
Proper investigation of the interplay between the dynamical interference effect discussed in the paper and the exited states
in the quantum dot is a open research topic beyond the scope of the present communication.
Nevertheless, we can make a simple estimate of the effect of finite $\Delta \varepsilon$
in the regime when backtunneling dominates ($\Gamma_c \ll \Delta_{\text{ptb}}$) using
the standard Markovian master equations~\cite{beenakker1PBI}.

We consider two levels $\varepsilon_1(t)$ and $\varepsilon_2(t)$, separated by a constant gap $\Delta \varepsilon$,
\begin{subequations} \label{eq:alleps2}
\begin{align}
  \varepsilon_1(t) & = \varepsilon_c +\Delta_{\rm ptb} (t-t_c)/\tau \, ,\\
  \varepsilon_2(t) & = \varepsilon_c +\Delta \varepsilon+ \Delta_{\rm ptb} (t-t_c)/\tau \, .
\end{align}
\end{subequations}
We assume equal tunnel broadenings for the levels, each equal to $\Gamma(t)$ as specified in Eq.~(4). 
In the limit of large charging energy, the kinetic equations~\cite{beenakker1PBI}
for the individual level occupation probabilities are ($i=1,2$):
\begin{multline} \label{eq:mastereq}
  \hbar \dot{P}_i(t)  =  \Gamma_i(t) 
  P_i(t) \bigl[1\!-\!f\bm{(}\varepsilon_i(t)\bm{)}\bigr] \\
  +  \Gamma_i(t)\bigl[1\!-n(t)\bigr] f\bm{(}\varepsilon_i(t)\bm{)} 
\end{multline}
with $n(t)=P_1(t)+P_2(t)$. An initial condition  at a sufficiently early time (we take $t_0=-20 \tau$) can be quantified by an effective excitation temperature $T^{\ast}$ (in energy units), $P_2(t_0)/P_1(t_0)= e^{-\Delta \varepsilon/T^{\ast}}$, $P_1(t_0)+P_2(t_0) =1$. No further mechanisms of excitation or relaxation
besides the charge tunneling [Eq.~\eqref{eq:mastereq}] are included in the model.

Solving \eqref{eq:mastereq} numerically for $\Delta \varepsilon > \Delta_{\rm ptb}$
results in the  same double-exponential shape for $n_f(\varepsilon_c)$ as for the single-level model, Eq.~(10c), except for an additional dip at $-\epsilon_c  \approx \Delta \varepsilon$.
The depth of the dip is proportional to the initial  fraction of the exited state, $P_2(t_0)$, see Fig.~\ref{fig:S1}(a).
For this reason we have interpreted the distance between the ground state and the first excited state features in
the charge capture probability measured in Fig.1(g) of Ref.~\onlinecite{kataoka2011}
as a direct indicator of $\Delta \varepsilon$
and thus obtained an estimate $\Delta \varepsilon / \Delta_{\text{ptb}}  \approx 4.5$ quoted in the main text.

We have verified that, on level of master equations \eqref{eq:mastereq}, the exponentially boosted derivative signal
$e^{-\varepsilon_c/\Delta_{\rm ptb}} \partial n_f/\partial \varepsilon_c$ (analogous to the one used in Fig.~\ref{fig:oscproof}
to demonstrate interference)
is not qualitatively affected by the presence of an excited state if $-\varepsilon_c < \Delta \varepsilon - 3 \Delta_{\text{ptb}}$,
see Fig.~\ref{fig:S1}(b). Since the smallest $|\varepsilon_c/\Delta_{\rm ptb}|$ for which the first interference minimum appear is $\approx 5$ 
[see Fig.~\ref{fig:ripples}(b)],
we may expect key results of our single-level modeling to hold for $\Delta \varepsilon \gtrsim 8 \Delta_{\rm ptb}$, as stated in the main text.


\begin{thebibliography}{36}
\expandafter\ifx\csname natexlab\endcsname\relax\def\natexlab#1{#1}\fi
\expandafter\ifx\csname bibnamefont\endcsname\relax
  \def\bibnamefont#1{#1}\fi
\expandafter\ifx\csname bibfnamefont\endcsname\relax
  \def\bibfnamefont#1{#1}\fi
\expandafter\ifx\csname citenamefont\endcsname\relax
  \def\citenamefont#1{#1}\fi
\expandafter\ifx\csname url\endcsname\relax
  \def\url#1{\texttt{#1}}\fi
\expandafter\ifx\csname urlprefix\endcsname\relax\def\urlprefix{URL }\fi
\providecommand{\bibinfo}[2]{#2}
\providecommand{\eprint}[2][]{\url{#2}}

\bibitem[{\citenamefont{Blumenthal et~al.}(2007)\citenamefont{Blumenthal,
  Kaestner, Li, Giblin, Janssen, Pepper, Anderson, Jones, and
  Ritchie}}]{blumenthal2007a}
\bibinfo{author}{\bibfnamefont{M.~D.} \bibnamefont{Blumenthal}},
  \bibinfo{author}{\bibfnamefont{B.}~\bibnamefont{Kaestner}},
  \bibinfo{author}{\bibfnamefont{L.}~\bibnamefont{Li}},
  \bibinfo{author}{\bibfnamefont{S.}~\bibnamefont{Giblin}},
  \bibinfo{author}{\bibfnamefont{T.~J. B.~M.} \bibnamefont{Janssen}},
  \bibinfo{author}{\bibfnamefont{M.}~\bibnamefont{Pepper}},
  \bibinfo{author}{\bibfnamefont{D.}~\bibnamefont{Anderson}},
  \bibinfo{author}{\bibfnamefont{G.}~\bibnamefont{Jones}}, \bibnamefont{and}
  \bibinfo{author}{\bibfnamefont{D.~A.} \bibnamefont{Ritchie}},
  \bibinfo{journal}{Nature Phys.} \textbf{\bibinfo{volume}{3}},
  \bibinfo{pages}{343 } (\bibinfo{year}{2007}).

\bibitem[{\citenamefont{{Giblin} et~al.}(2012)\citenamefont{{Giblin},
  {Kataoka}, {Fletcher}, {See}, {Janssen}, {Griffiths}, {Jones}, {Farrer}, and
  {Ritchie}}}]{Giblin2012}
\bibinfo{author}{\bibfnamefont{S.~P.} \bibnamefont{{Giblin}}},
  \bibinfo{author}{\bibfnamefont{M.}~\bibnamefont{{Kataoka}}},
  \bibinfo{author}{\bibfnamefont{J.~D.} \bibnamefont{{Fletcher}}},
  \bibinfo{author}{\bibfnamefont{P.}~\bibnamefont{{See}}},
  \bibinfo{author}{\bibfnamefont{T.~J.~B.~M.} \bibnamefont{{Janssen}}},
  \bibinfo{author}{\bibfnamefont{J.~P.} \bibnamefont{{Griffiths}}},
  \bibinfo{author}{\bibfnamefont{G.~A.~C.} \bibnamefont{{Jones}}},
  \bibinfo{author}{\bibfnamefont{I.}~\bibnamefont{{Farrer}}}, \bibnamefont{and}
  \bibinfo{author}{\bibfnamefont{D.~A.} \bibnamefont{{Ritchie}}},
  \bibinfo{journal}{Nat. Commun.},  \textbf{\bibinfo{volume}{3}},
  \bibinfo{pages}{930} (\bibinfo{year}{2012}).


\bibitem[{\citenamefont{Leicht et~al.}(2011)\citenamefont{Leicht, Mirovsky,
  Kaestner, Hohls, Kashcheyevs, Kurganova, Zeitler, Weimann, Pierz, and
  Schumacher}}]{kaestner2010d}
\bibinfo{author}{\bibfnamefont{C.}~\bibnamefont{Leicht}},
  \bibinfo{author}{\bibfnamefont{P.}~\bibnamefont{Mirovsky}},
  \bibinfo{author}{\bibfnamefont{B.}~\bibnamefont{Kaestner}},
  \bibinfo{author}{\bibfnamefont{F.}~\bibnamefont{Hohls}},
  \bibinfo{author}{\bibfnamefont{V.}~\bibnamefont{Kashcheyevs}},
  \bibinfo{author}{\bibfnamefont{E.~V.} \bibnamefont{Kurganova}},
  \bibinfo{author}{\bibfnamefont{U.}~\bibnamefont{Zeitler}},
  \bibinfo{author}{\bibfnamefont{T.}~\bibnamefont{Weimann}},
  \bibinfo{author}{\bibfnamefont{K.}~\bibnamefont{Pierz}}, \bibnamefont{and}
  \bibinfo{author}{\bibfnamefont{H.~W.} \bibnamefont{Schumacher}},
  \bibinfo{journal}{Semicond. Sci. Technol.} \textbf{\bibinfo{volume}{26}},
  \bibinfo{pages}{055010} (\bibinfo{year}{2011}).



\bibitem[{\citenamefont{F\`eve et~al.}(2007)\citenamefont{F\`eve, Mah\'e,
  Berroir, Kontos, Pla\ifmmode~\mbox{\c{c}}\else \c{c}\fi{}ais, Glattli,
  Cavanna, Etienne, and Jin}}]{Feve2007}
\bibinfo{author}{\bibfnamefont{G.}~\bibnamefont{F\`eve}},
  \bibinfo{author}{\bibfnamefont{A.}~\bibnamefont{Mah\'e}},
  \bibinfo{author}{\bibfnamefont{J.-M.} \bibnamefont{Berroir}},
  \bibinfo{author}{\bibfnamefont{T.}~\bibnamefont{Kontos}},
  \bibinfo{author}{\bibfnamefont{B.}~\bibnamefont{Pla\ifmmode~\mbox{\c{c}}\else
  \c{c}\fi{}ais}}, \bibinfo{author}{\bibfnamefont{D.~C.}
  \bibnamefont{Glattli}},
  \bibinfo{author}{\bibfnamefont{A.}~\bibnamefont{Cavanna}},
  \bibinfo{author}{\bibfnamefont{B.}~\bibnamefont{Etienne}}, \bibnamefont{and}
  \bibinfo{author}{\bibfnamefont{Y.}~\bibnamefont{Jin}},
  \bibinfo{journal}{Science} \textbf{\bibinfo{volume}{316}},
  \bibinfo{pages}{1169} (\bibinfo{year}{2007}).

\bibitem[{\citenamefont{{Bocquillon} et~al.}(2012)\citenamefont{{Bocquillon},
  {Parmentier}, {Grenier}, {Berroir}, {Degiovanni}, {Glattli}, {Pla{\c c}ais},
  {Cavanna}, {Jin}, and {F{\`e}ve}}}]{Feve2012exp}
\bibinfo{author}{\bibfnamefont{E.}~\bibnamefont{{Bocquillon}}},
  \bibinfo{author}{\bibfnamefont{F.~D.} \bibnamefont{{Parmentier}}},
  \bibinfo{author}{\bibfnamefont{C.}~\bibnamefont{{Grenier}}},
  \bibinfo{author}{\bibfnamefont{J.-M.} \bibnamefont{{Berroir}}},
  \bibinfo{author}{\bibfnamefont{P.}~\bibnamefont{{Degiovanni}}},
  \bibinfo{author}{\bibfnamefont{D.~C.} \bibnamefont{{Glattli}}},
  \bibinfo{author}{\bibfnamefont{B.}~\bibnamefont{{Pla{\c c}ais}}},
  \bibinfo{author}{\bibfnamefont{A.}~\bibnamefont{{Cavanna}}},
  \bibinfo{author}{\bibfnamefont{Y.}~\bibnamefont{{Jin}}}, \bibnamefont{and}
  \bibinfo{author}{\bibfnamefont{G.}~\bibnamefont{{F{\`e}ve}}},
  \bibinfo{journal}{Phys. Rev. Lett.} \textbf{\bibinfo{volume}{108}},
  \bibinfo{pages}{196803} (\bibinfo{year}{2012}).

\bibitem[{\citenamefont{Pan et~al.}(2012)\citenamefont{Pan, Chen, Lu,
  Weinfurter, Zeilinger, and \ifmmode~\dot{Z}\else
  \.{Z}\fi{}ukowski}}]{Zeilinger2012}
\bibinfo{author}{\bibfnamefont{J.-W.} \bibnamefont{Pan}},
  \bibinfo{author}{\bibfnamefont{Z.-B.} \bibnamefont{Chen}},
  \bibinfo{author}{\bibfnamefont{C.-Y.} \bibnamefont{Lu}},
  \bibinfo{author}{\bibfnamefont{H.}~\bibnamefont{Weinfurter}},
  \bibinfo{author}{\bibfnamefont{A.}~\bibnamefont{Zeilinger}},
  \bibnamefont{and}
  \bibinfo{author}{\bibfnamefont{M.}~\bibnamefont{\ifmmode~\dot{Z}\else
  \.{Z}\fi{}ukowski}}, \bibinfo{journal}{Rev. Mod. Phys.}
  \textbf{\bibinfo{volume}{84}}, \bibinfo{pages}{777} (\bibinfo{year}{2012}).

\bibitem[{\citenamefont{Bertoni et~al.}(2000)\citenamefont{Bertoni, Bordone,
  Brunetti, Jacoboni, and Reggiani}}]{Bertoni2000}
\bibinfo{author}{\bibfnamefont{A.}~\bibnamefont{Bertoni}},
  \bibinfo{author}{\bibfnamefont{P.}~\bibnamefont{Bordone}},
  \bibinfo{author}{\bibfnamefont{R.}~\bibnamefont{Brunetti}},
  \bibinfo{author}{\bibfnamefont{C.}~\bibnamefont{Jacoboni}}, \bibnamefont{and}
  \bibinfo{author}{\bibfnamefont{S.}~\bibnamefont{Reggiani}},
  \bibinfo{journal}{Phys. Rev. Lett.} \textbf{\bibinfo{volume}{84}},
  \bibinfo{pages}{5912} (\bibinfo{year}{2000}).

\bibitem[{\citenamefont{Ol'khovskaya et~al.}(2008)\citenamefont{Ol'khovskaya,
  Splettstoesser, Moskalets, and B\"uttiker}}]{Olkhovskaya2008}
\bibinfo{author}{\bibfnamefont{S.}~\bibnamefont{Ol'khovskaya}},
  \bibinfo{author}{\bibfnamefont{J.}~\bibnamefont{Splettstoesser}},
  \bibinfo{author}{\bibfnamefont{M.}~\bibnamefont{Moskalets}},
  \bibnamefont{and}
  \bibinfo{author}{\bibfnamefont{M.}~\bibnamefont{B\"uttiker}},
  \bibinfo{journal}{Phys. Rev. Lett.} \textbf{\bibinfo{volume}{101}},
  \bibinfo{pages}{166802} (\bibinfo{year}{2008}).

\bibitem[{\citenamefont{Splettstoesser
  et~al.}(2009)\citenamefont{Splettstoesser, Moskalets, and
  B\"uttiker}}]{SplettstoesserInterf2009}
\bibinfo{author}{\bibfnamefont{J.}~\bibnamefont{Splettstoesser}},
  \bibinfo{author}{\bibfnamefont{M.}~\bibnamefont{Moskalets}},
  \bibnamefont{and}
  \bibinfo{author}{\bibfnamefont{M.}~\bibnamefont{B\"uttiker}},
  \bibinfo{journal}{Phys. Rev. Lett.} \textbf{\bibinfo{volume}{103}},
  \bibinfo{pages}{076804} (\bibinfo{year}{2009}).

\bibitem[{\citenamefont{Haack et~al.}(2011)\citenamefont{Haack, Moskalets,
  Splettstoesser, and B\"uttiker}}]{Haack2012}
\bibinfo{author}{\bibfnamefont{G.}~\bibnamefont{Haack}},
  \bibinfo{author}{\bibfnamefont{M.}~\bibnamefont{Moskalets}},
  \bibinfo{author}{\bibfnamefont{J.}~\bibnamefont{Splettstoesser}},
  \bibnamefont{and}
  \bibinfo{author}{\bibfnamefont{M.}~\bibnamefont{B\"uttiker}},
  \bibinfo{journal}{Phys. Rev. B} \textbf{\bibinfo{volume}{84}},
  \bibinfo{pages}{081303R} (\bibinfo{year}{2011}).

\bibitem[{\citenamefont{Zimmerman and Keller}(2003)}]{Zimmerman2003}
\bibinfo{author}{\bibfnamefont{N.~M.} \bibnamefont{Zimmerman}}
  \bibnamefont{and} \bibinfo{author}{\bibfnamefont{M.~W.}
  \bibnamefont{Keller}}, \bibinfo{journal}{Meas. Sci. Technol.}
  \textbf{\bibinfo{volume}{14}}, \bibinfo{pages}{1237} (\bibinfo{year}{2003}).

\bibitem[{\citenamefont{Milton et~al.}(2010)\citenamefont{Milton, Williams, and
  Forbes}}]{Milton2010}
\bibinfo{author}{\bibfnamefont{M.~J.~T.} \bibnamefont{Milton}},
  \bibinfo{author}{\bibfnamefont{J.~M.} \bibnamefont{Williams}},
  \bibnamefont{and} \bibinfo{author}{\bibfnamefont{A.~B.}
  \bibnamefont{Forbes}}, \bibinfo{journal}{Metrologia}
  \textbf{\bibinfo{volume}{47}}, \bibinfo{pages}{279} (\bibinfo{year}{2010}).

\bibitem[{\citenamefont{Brandes}(2010)}]{Brandes2010}
\bibinfo{author}{\bibfnamefont{T.}~\bibnamefont{Brandes}},
  \bibinfo{journal}{Phys. Rev. Lett.} \textbf{\bibinfo{volume}{105}},
  \bibinfo{pages}{060602} (\bibinfo{year}{2010}).

\bibitem[{\citenamefont{Fricke et~al.}(2011)\citenamefont{Fricke, Hohls,
  Ubbelohde, Kaestner, Kashcheyevs, Leicht, Mirovsky, Pierz, Schumacher, and
  Haug}}]{fricke2011}
\bibinfo{author}{\bibfnamefont{L.}~\bibnamefont{Fricke}},
  \bibinfo{author}{\bibfnamefont{F.}~\bibnamefont{Hohls}},
  \bibinfo{author}{\bibfnamefont{N.}~\bibnamefont{Ubbelohde}},
  \bibinfo{author}{\bibfnamefont{B.}~\bibnamefont{Kaestner}},
  \bibinfo{author}{\bibfnamefont{V.}~\bibnamefont{Kashcheyevs}},
  \bibinfo{author}{\bibfnamefont{C.}~\bibnamefont{Leicht}},
  \bibinfo{author}{\bibfnamefont{P.}~\bibnamefont{Mirovsky}},
  \bibinfo{author}{\bibfnamefont{K.}~\bibnamefont{Pierz}},
  \bibinfo{author}{\bibfnamefont{H.~W.} \bibnamefont{Schumacher}},
  \bibnamefont{and} \bibinfo{author}{\bibfnamefont{R.~J.} \bibnamefont{Haug}},
  \bibinfo{journal}{Phys. Rev. B} \textbf{\bibinfo{volume}{83}},
  \bibinfo{pages}{193306} (\bibinfo{year}{2011}).

\bibitem[{\citenamefont{Liu and Niu}(1993)}]{liu1993}
\bibinfo{author}{\bibfnamefont{C.}~\bibnamefont{Liu}} \bibnamefont{and}
  \bibinfo{author}{\bibfnamefont{Q.}~\bibnamefont{Niu}},
  \bibinfo{journal}{Phys. Rev. B} \textbf{\bibinfo{volume}{47}},
  \bibinfo{pages}{13031} (\bibinfo{year}{1993}).

\bibitem[{\citenamefont{Flensberg et~al.}(1999)\citenamefont{Flensberg, Niu,
  and Pustilnik}}]{Flensberg1999}
\bibinfo{author}{\bibfnamefont{K.}~\bibnamefont{Flensberg}},
  \bibinfo{author}{\bibfnamefont{Q.}~\bibnamefont{Niu}}, \bibnamefont{and}
  \bibinfo{author}{\bibfnamefont{M.}~\bibnamefont{Pustilnik}},
  \bibinfo{journal}{Phys. Rev. B} \textbf{\bibinfo{volume}{60}},
  \bibinfo{pages}{R16291} (\bibinfo{year}{1999}).


\bibitem[{\citenamefont{Kataoka et~al.}(2011)\citenamefont{Kataoka, Fletcher,
  See, Giblin, Janssen, Griffiths, Jones, Farrer, and Ritchie}}]{kataoka2011}
\bibinfo{author}{\bibfnamefont{M.}~\bibnamefont{Kataoka}},
  \bibinfo{author}{\bibfnamefont{J.~D.} \bibnamefont{Fletcher}},
  \bibinfo{author}{\bibfnamefont{P.}~\bibnamefont{See}},
  \bibinfo{author}{\bibfnamefont{S.~P.} \bibnamefont{Giblin}},
  \bibinfo{author}{\bibfnamefont{T.~J. B.~M.} \bibnamefont{Janssen}},
  \bibinfo{author}{\bibfnamefont{J.~P.} \bibnamefont{Griffiths}},
  \bibinfo{author}{\bibfnamefont{G.~A.~C.} \bibnamefont{Jones}},
  \bibinfo{author}{\bibfnamefont{I.}~\bibnamefont{Farrer}}, \bibnamefont{and}
  \bibinfo{author}{\bibfnamefont{D.~A.} \bibnamefont{Ritchie}},
  \bibinfo{journal}{Phys. Rev. Lett.} \textbf{\bibinfo{volume}{106}},
  \bibinfo{pages}{126801} (\bibinfo{year}{2011}).


\bibitem[{\citenamefont{Aiz\v{\i}n et~al.}(1998)\citenamefont{Aiz\v{\i}n,
  Gumbs, and Pepper}}]{Aizin1998}
\bibinfo{author}{\bibfnamefont{G.~R.} \bibnamefont{Aiz\v{\i}n}},
  \bibinfo{author}{\bibfnamefont{G.}~\bibnamefont{Gumbs}}, \bibnamefont{and}
  \bibinfo{author}{\bibfnamefont{M.}~\bibnamefont{Pepper}},
  \bibinfo{journal}{Phys. Rev. B} \textbf{\bibinfo{volume}{58}},
  \bibinfo{pages}{10589} (\bibinfo{year}{1998}).

\bibitem[{\citenamefont{Fujiwara et~al.}(2008)\citenamefont{Fujiwara,
  Nishiguchi, and Ono}}]{fujiwara2008}
\bibinfo{author}{\bibfnamefont{A.}~\bibnamefont{Fujiwara}},
  \bibinfo{author}{\bibfnamefont{K.}~\bibnamefont{Nishiguchi}},
  \bibnamefont{and} \bibinfo{author}{\bibfnamefont{Y.}~\bibnamefont{Ono}},
  \bibinfo{journal}{Appl. Phys. Lett.} \textbf{\bibinfo{volume}{92}},
  \bibinfo{pages}{042102} (\bibinfo{year}{2008}).

\bibitem[{\citenamefont{Kaestner et~al.}(2009)\citenamefont{Kaestner, Leicht,
  Kashcheyevs, Pierz, Siegner, and Schumacher}}]{kaestner2009a}
\bibinfo{author}{\bibfnamefont{B.}~\bibnamefont{Kaestner}},
  \bibinfo{author}{\bibfnamefont{C.}~\bibnamefont{Leicht}},
  \bibinfo{author}{\bibfnamefont{V.}~\bibnamefont{Kashcheyevs}},
  \bibinfo{author}{\bibfnamefont{K.}~\bibnamefont{Pierz}},
  \bibinfo{author}{\bibfnamefont{U.}~\bibnamefont{Siegner}}, \bibnamefont{and}
  \bibinfo{author}{\bibfnamefont{H.~W.} \bibnamefont{Schumacher}},
  \bibinfo{journal}{Appl. Phys. Lett.} \textbf{\bibinfo{volume}{94}},
  \bibinfo{pages}{012106} (\bibinfo{year}{2009}).

\bibitem[{\citenamefont{Kashcheyevs and Kaestner}(2010)}]{kaestner2010a}
\bibinfo{author}{\bibfnamefont{V.}~\bibnamefont{Kashcheyevs}} \bibnamefont{and}
  \bibinfo{author}{\bibfnamefont{B.}~\bibnamefont{Kaestner}},
  \bibinfo{journal}{Phys. Rev. Lett.} \textbf{\bibinfo{volume}{104}},
  \bibinfo{pages}{186805} (\bibinfo{year}{2010}).

\bibitem[{\citenamefont{Keeling et~al.}(2008)\citenamefont{Keeling, Shytov, and
  Levitov}}]{Keeling2008}
\bibinfo{author}{\bibfnamefont{J.}~\bibnamefont{Keeling}},
  \bibinfo{author}{\bibfnamefont{A.~V.} \bibnamefont{Shytov}},
  \bibnamefont{and} \bibinfo{author}{\bibfnamefont{L.~S.}
  \bibnamefont{Levitov}}, \bibinfo{journal}{Phys. Rev. Lett.}
  \textbf{\bibinfo{volume}{101}}, \bibinfo{pages}{196404}
  (\bibinfo{year}{2008}).

\bibitem[{\citenamefont{Parmentier et~al.}(2012)\citenamefont{Parmentier,
  Bocquillon, Berroir, Glattli, Pla\ifmmode~\mbox{\c{c}}\else \c{c}\fi{}ais,
  F\`eve, Albert, Flindt, and B\"uttiker}}]{Parmentier2012}
\bibinfo{author}{\bibfnamefont{F.~D.} \bibnamefont{Parmentier}},
  \bibinfo{author}{\bibfnamefont{E.}~\bibnamefont{Bocquillon}},
  \bibinfo{author}{\bibfnamefont{J.-M.} \bibnamefont{Berroir}},
  \bibinfo{author}{\bibfnamefont{D.~C.} \bibnamefont{Glattli}},
  \bibinfo{author}{\bibfnamefont{B.}~\bibnamefont{Pla\ifmmode~\mbox{\c{c}}\else
  \c{c}\fi{}ais}}, \bibinfo{author}{\bibfnamefont{G.}~\bibnamefont{F\`eve}},
  \bibinfo{author}{\bibfnamefont{M.}~\bibnamefont{Albert}},
  \bibinfo{author}{\bibfnamefont{C.}~\bibnamefont{Flindt}}, \bibnamefont{and}
  \bibinfo{author}{\bibfnamefont{M.}~\bibnamefont{B\"uttiker}},
  \bibinfo{journal}{Phys. Rev. B} \textbf{\bibinfo{volume}{85}},
  \bibinfo{pages}{165438} (\bibinfo{year}{2012}).

\bibitem[{\citenamefont{Jauho et~al.}(1994)\citenamefont{Jauho, Wingreen, and
  Meir}}]{Jauho1994}
\bibinfo{author}{\bibfnamefont{A.-P.} \bibnamefont{Jauho}},
  \bibinfo{author}{\bibfnamefont{N.~S.} \bibnamefont{Wingreen}},
  \bibnamefont{and} \bibinfo{author}{\bibfnamefont{Y.}~\bibnamefont{Meir}},
  \bibinfo{journal}{Phys. Rev. B} \textbf{\bibinfo{volume}{50}},
  \bibinfo{pages}{5528} (\bibinfo{year}{1994}).

\bibitem[{\citenamefont{Landau and Lifshitz}(1981)}]{LandauQMlz}
\bibinfo{author}{\bibfnamefont{L.~D.} \bibnamefont{Landau}} \bibnamefont{and}
  \bibinfo{author}{\bibfnamefont{E.~M.} \bibnamefont{Lifshitz}},
  \emph{\bibinfo{title}{Quantum Mechanics}}
  (\bibinfo{publisher}{Butterworth-Heinemann}, \bibinfo{address}{Amsterdam}, \bibinfo{year}{1981}),
  vol.~\bibinfo{volume}{3}, pp. \bibinfo{pages}{195--198}.


\bibitem[{\citenamefont{Oliver et~al.}(2005)\citenamefont{Oliver, Yu, Lee,
  Berggren, Levitov, and Orlando}}]{LevitovScience2005}
\bibinfo{author}{\bibfnamefont{W.~D.} \bibnamefont{Oliver}},
  \bibinfo{author}{\bibfnamefont{Y.}~\bibnamefont{Yu}},
  \bibinfo{author}{\bibfnamefont{J.~C.} \bibnamefont{Lee}},
  \bibinfo{author}{\bibfnamefont{K.~K.} \bibnamefont{Berggren}},
  \bibinfo{author}{\bibfnamefont{L.~S.} \bibnamefont{Levitov}},
  \bibnamefont{and} \bibinfo{author}{\bibfnamefont{T.~P.}
  \bibnamefont{Orlando}}, \bibinfo{journal}{Science}
  \textbf{\bibinfo{volume}{310}}, \bibinfo{pages}{1653} (\bibinfo{year}{2005}).

\bibitem[{\citenamefont{Mark et~al.}(2007)\citenamefont{Mark, Kraemer,
  Waldburger, Herbig, Chin, N\"agerl, and Grimm}}]{Kraemer2007}
\bibinfo{author}{\bibfnamefont{M.}~\bibnamefont{Mark}},
  \bibinfo{author}{\bibfnamefont{T.}~\bibnamefont{Kraemer}},
  \bibinfo{author}{\bibfnamefont{P.}~\bibnamefont{Waldburger}},
  \bibinfo{author}{\bibfnamefont{J.}~\bibnamefont{Herbig}},
  \bibinfo{author}{\bibfnamefont{C.}~\bibnamefont{Chin}},
  \bibinfo{author}{\bibfnamefont{H.-C.} \bibnamefont{N\"agerl}},
  \bibnamefont{and} \bibinfo{author}{\bibfnamefont{R.}~\bibnamefont{Grimm}},
  \bibinfo{journal}{Phys. Rev. Lett.} \textbf{\bibinfo{volume}{99}},
  \bibinfo{pages}{113201} (\bibinfo{year}{2007}).

\bibitem[{\citenamefont{Shevchenko et~al.}(2010)\citenamefont{Shevchenko,
  Ashhab, and Nori}}]{Shevchenko2010}
\bibinfo{author}{\bibfnamefont{S.}~\bibnamefont{Shevchenko}},
  \bibinfo{author}{\bibfnamefont{S.}~\bibnamefont{Ashhab}}, \bibnamefont{and}
  \bibinfo{author}{\bibfnamefont{F.}~\bibnamefont{Nori}},
  \bibinfo{journal}{Phys. Rep.} \textbf{\bibinfo{volume}{492}},
  \bibinfo{pages}{1} (\bibinfo{year}{2010}).

\bibitem[{\citenamefont{Moskalets and B\"uttiker}(2002)}]{moskalets2002B}
\bibinfo{author}{\bibfnamefont{M.}~\bibnamefont{Moskalets}} \bibnamefont{and}
  \bibinfo{author}{\bibfnamefont{M.}~\bibnamefont{B\"uttiker}},
  \bibinfo{journal}{Phys. Rev. B} \textbf{\bibinfo{volume}{66}},
  \bibinfo{pages}{205320} (\bibinfo{year}{2002}).

\bibitem[{\citenamefont{Kowenhven}(2002)}]{kouwenhoven2001}
\bibinfo{author}{\bibfnamefont{L.~P.}~\bibnamefont{Kouwenhoven}},
\bibinfo{author}{\bibfnamefont{D.~G.}~\bibnamefont{Austing}},
\bibnamefont{and} \bibinfo{author}{\bibfnamefont{S.}~\bibnamefont{Tarucha}},
   \bibinfo{journal}{Rep. Prog. Phys.} \textbf{\bibinfo{volume}{64}},
  \bibinfo{pages}{701} (\bibinfo{year}{2001}).


\bibitem[{\citenamefont{Hanggi}(2006)}]{hanggiShow}
  \bibinfo{author}{\bibfnamefont{S.}~\bibnamefont{Kohler}},
  \bibinfo{author}{\bibfnamefont{J.}~\bibnamefont{Lehmann}},
  \bibnamefont{and}
\bibinfo{author}{\bibfnamefont{P.}~\bibnamefont{H\"{a}nggi}},
\bibinfo{journal}{Phys. Rep.} \textbf{\bibinfo{volume}{406}},
  \bibinfo{pages}{379} (\bibinfo{year}{2005}).

\bibitem[{\citenamefont{Arrachea and Moskalets}(2006)}]{Arrachea2006}
\bibinfo{author}{\bibfnamefont{L.}~\bibnamefont{Arrachea}} \bibnamefont{and}
  \bibinfo{author}{\bibfnamefont{M.}~\bibnamefont{Moskalets}},
  \bibinfo{journal}{Phys. Rev. B} \textbf{\bibinfo{volume}{74}},
  \bibinfo{pages}{245322} (\bibinfo{year}{2006}).

\bibitem[{\citenamefont{Hanggi}(xxx)}]{hanggiFloquet}
\bibinfo{author}{\bibfnamefont{M.}~\bibnamefont{Grifoni}} \bibnamefont{and}
  \bibinfo{author}{\bibfnamefont{P.}~\bibnamefont{H\"{a}nggi}},
  \bibinfo{journal}{Phys. Rep.} \textbf{\bibinfo{volume}{304}},
  \bibinfo{pages}{229} (\bibinfo{year}{1998}).

\bibitem[{\citenamefont{Moskalets et~al.}(2008)\citenamefont{Moskalets,
  Samuelsson, and B\"uttiker}}]{Floquet2008prl}
\bibinfo{author}{\bibfnamefont{M.}~\bibnamefont{Moskalets}},
  \bibinfo{author}{\bibfnamefont{P.}~\bibnamefont{Samuelsson}},
  \bibnamefont{and}
  \bibinfo{author}{\bibfnamefont{M.}~\bibnamefont{B\"uttiker}},
  \bibinfo{journal}{Phys. Rev. Lett.} \textbf{\bibinfo{volume}{100}},
  \bibinfo{pages}{086601} (\bibinfo{year}{2008}).

\bibitem[{\citenamefont{battista2011}(2011)\citenamefont{Moskalets,
  Samuelsson, and B\"uttiker}}]{battista2011}
\bibinfo{author}{\bibfnamefont{F.}~\bibnamefont{Battista}} \bibnamefont{and}
  \bibinfo{author}{\bibfnamefont{P.}~\bibnamefont{Samuelsson}},
  \bibinfo{journal}{Phys. Rev. B} \textbf{\bibinfo{volume}{83}},
  \bibinfo{pages}{125324} (\bibinfo{year}{2011}).


\bibitem[{\citenamefont{Kaestner et~al.}(2008)\citenamefont{Kaestner,
  Kashcheyevs, Amakawa, Blumenthal, Li, Janssen, Hein, Pierz, Weimann, Siegner
  et~al.}}]{Kaestner2007c}
\bibinfo{author}{\bibfnamefont{B.}~\bibnamefont{Kaestner}},
  \bibinfo{author}{\bibfnamefont{V.}~\bibnamefont{Kashcheyevs}},
  \bibinfo{author}{\bibfnamefont{S.}~\bibnamefont{Amakawa}},
  \bibinfo{author}{\bibfnamefont{M.~D.} \bibnamefont{Blumenthal}},
  \bibinfo{author}{\bibfnamefont{L.}~\bibnamefont{Li}},
  \bibinfo{author}{\bibfnamefont{T.~J. B.~M.} \bibnamefont{Janssen}},
  \bibinfo{author}{\bibfnamefont{G.}~\bibnamefont{Hein}},
  \bibinfo{author}{\bibfnamefont{K.}~\bibnamefont{Pierz}},
  \bibinfo{author}{\bibfnamefont{T.}~\bibnamefont{Weimann}},
  \bibinfo{author}{\bibfnamefont{U.}~\bibnamefont{Siegner}},
  \bibnamefont{et~al.}, \bibinfo{journal}{Phys. Rev. B}
  \textbf{\bibinfo{volume}{77}}, \bibinfo{pages}{153301}
  (\bibinfo{year}{2008}).

\bibitem[{\citenamefont{Giblin et~al.}(2010)\citenamefont{Giblin, Wright,
  Fletcher, Kataoka, Pepper, Janssen, Ritchie, Nicoll, Anderson, and
  Jones}}]{Giblin2010}
\bibinfo{author}{\bibfnamefont{S.~P.} \bibnamefont{Giblin}},
  \bibinfo{author}{\bibfnamefont{S.~J.} \bibnamefont{Wright}},
  \bibinfo{author}{\bibfnamefont{J.~D.} \bibnamefont{Fletcher}},
  \bibinfo{author}{\bibfnamefont{M.}~\bibnamefont{Kataoka}},
  \bibinfo{author}{\bibfnamefont{M.}~\bibnamefont{Pepper}},
  \bibinfo{author}{\bibfnamefont{T.~J. B.~M.} \bibnamefont{Janssen}},
  \bibinfo{author}{\bibfnamefont{D.~A.} \bibnamefont{Ritchie}},
  \bibinfo{author}{\bibfnamefont{C.~A.} \bibnamefont{Nicoll}},
  \bibinfo{author}{\bibfnamefont{D.}~\bibnamefont{Anderson}}, \bibnamefont{and}
  \bibinfo{author}{\bibfnamefont{G.~A.~C.} \bibnamefont{Jones}},
  \bibinfo{journal}{New J. Phys.} \textbf{\bibinfo{volume}{12}},
  \bibinfo{pages}{073013} (\bibinfo{year}{2010}).


\bibitem[{\citenamefont{Yamahata et~al.}(2011)\citenamefont{Yamahata,
  Nishiguchi, and Fujiwara}}]{yamahata2011}
\bibinfo{author}{\bibfnamefont{G.}~\bibnamefont{Yamahata}},
  \bibinfo{author}{\bibfnamefont{K.}~\bibnamefont{Nishiguchi}},
  \bibnamefont{and} \bibinfo{author}{\bibfnamefont{A.}~\bibnamefont{Fujiwara}},
  \bibinfo{journal}{App. Phys. Lett.} \textbf{\bibinfo{volume}{98}},
  \bibinfo{pages}{222104} (\bibinfo{year}{2011}).

\bibitem[{\citenamefont{{Ya. Azbel'}}(1992)}]{Azbel1992}
\bibinfo{author}{\bibfnamefont{M.}~\bibnamefont{{Ya. Azbel'}}},
  \bibinfo{journal}{Europhys. Lett.} \textbf{\bibinfo{volume}{18}},
  \bibinfo{pages}{537} (\bibinfo{year}{1992}).

\bibitem[{Note2()}]{Note2}
\bibinfo{note}{$H_{12} \propto V^2(t)$ since it is a second order
  process for electrons in the lead}.

\bibitem[{\citenamefont{{Beenakker}}(1991)}]{beenakker1PBI}
\bibinfo{author}{\bibfnamefont{C. W. J.}~\bibnamefont{{Beenakker}}},
  \bibinfo{journal}{Phys. Rev. B} \textbf{\bibinfo{volume}{44}},
  \bibinfo{pages}{1646} (\bibinfo{year}{1991}).

\end{thebibliography}

\end{document}